\documentclass[journal,11pt,final,a4paper,twocolumn,oneside]{IEEEtran} \pdfoutput=1

\usepackage[utf8]{inputenc} \usepackage[T1]{fontenc} \usepackage[hidelinks]{hyperref} \usepackage{subfig} \usepackage{graphicx} \usepackage{amsmath,amssymb,stmaryrd} \usepackage{mathtools} \usepackage{bm} \DeclareMathVersion{bold} \interdisplaylinepenalty=2500 \usepackage{algorithm} \usepackage[noend]{algorithmic} \algsetup{linenodelimiter=\ } \algsetup{linenosize=\scriptsize} \algsetup{indent=1.5em} \usepackage[font=footnotesize]{caption} \usepackage{fixltx2e} \usepackage{dblfloatfix} \usepackage{url}

\makeatletter \begingroup \catcode`\_=\active \protected\gdef_{\@ifnextchar|\subtextup\sb} \endgroup \def\subtextup|#1|{\sb{\textup{#1}}} \AtBeginDocument{\catcode`\_=12 \mathcode`\_=32768 } \newcommand{\nullindent}{ \@afterheading \@afterindentfalse } \let\oldalign=\align \let\endoldalign=\endalign \renewenvironment{align}{\vspace{-5.0pt}\oldalign}{\endoldalign\par\aftergroup\nullindent\vspace{10.0pt}} \renewenvironment{align*}{ \vspace{-5.0pt} \start@align\@ne\st@rredtrue\m@ne }{ \endalign } \makeatother \usepackage{fancyhdr}   \usepackage{siunitx}           \DeclareSIUnit\bpp{bpp} \newcommand{\g}{\cdot} \newcommand{\abs}[1]{\left|{#1}\right|} \renewcommand{\v}[1]{\mathbf{#1}} \newcommand{\m}[1]{\mathbf{#1}} \newcommand{\T}{^{\mathrm{T}}} \newcommand{\MT}{^{-\mathrm{T}}} \newcommand{\algarrow}{\,\leftarrow\,} \renewcommand{\exp}[1]{\text{exp}\!\left(#1\right)}  \makeatletter \newcommand*{\rom}[1]{\expandafter\@slowromancap\romannumeral #1@} \makeatother

\begin{document} \title{Improving Smoothed $\ell_0$ Norm in Compressive Sensing Using Adaptive Parameter Selection} \author{Christian Schou Oxvig, Patrick Steffen Pedersen, Thomas Arildsen, and Torben Larsen \thanks{This work was partially financed by The Danish Council for Strategic Research under grant number 09-067056.}} \maketitle

\setlength{\headheight}{80pt} \thispagestyle{fancy} \lhead{\footnotesize This is an early draft of the published paper ``SURPASSING THE THEORETICAL 1-NORM PHASE TRANSITION IN COMPRESSIVE SENSING BY TUNING THE SMOOTHED L0 ALGORITHM'' (published in the IEEE ICASSP 2013). The published paper differs significantly from this edition. However, parts of this paper are subject to the following copyright notice: \\ Copyright 2013 IEEE. Published in the IEEE 2013 International Conference on Acoustics, Speech, and Signal Processing (ICASSP 2013), scheduled for 26-31 May 2013 in Vancouver, British Columbia, Canada. Personal use of this material is permitted. However, permission to reprint/republish this material for advertising or promotional purposes or for creating new collective works for resale or redistribution to servers or lists, or to reuse any copyrighted component of this work in other works, must be obtained from the IEEE. Contact: Manager, Copyrights and Permissions / IEEE Service Center / 445 Hoes Lane / P.O. Box 1331 / Piscataway, NJ 08855-1331, USA. Telephone: + Intl. 908-562-3966.}

\begin{abstract} Signal reconstruction in compressive sensing involves finding a sparse solution that satisfies a set of linear constraints. Several approaches to this problem have been considered in existing reconstruction algorithms. They each provide a trade-off between reconstruction capabilities and required computation time. In an attempt to push the limits for this trade-off, we consider a smoothed $\ell_0$ norm (SL0) algorithm in a noiseless setup. We argue that using a set of carefully chosen parameters in our proposed adaptive SL0 algorithm may result in significantly better reconstruction capabilities in terms of phase transition while retaining the same required computation time as existing SL0 algorithms. A large set of simulations further support this claim. Simulations even reveal that the theoretical $\ell_1$ curve may be surpassed in major parts of the phase space.

\end{abstract}

\section{Introduction} \label{sec:intro} \IEEEPARstart{T}{he} Compressive Sensing (CS) signal acquisition paradigm asserts that one can successfully recover certain signals sampled far below their Nyquist frequencies given they are sparse in some dictionary \cite{candes2008}. The Fourier dictionary for frequency sparse signals is an example of this. Encouraged by this assertion, the usual sample and then compress setup can be combined into a single efficient step. Signals acquired in this fashion do, however, have to be reconstructed which, in the noiseless case, entails a non-convex optimisation problem of the form:

\begin{align} \begin{aligned} &\text{minimise} && ||\hat{\v{x}}||_0 \\ &\text{subject to} && \v{y} = \m{A} \hat{\v{x}} \end{aligned} \label{eq:compressive_sensing-general_problem} \end{align}

\noindent where $\hat{\v{x}} \in \mathbb{R}^{N \times 1}$ is the reconstructed signal, $\m{A} \in \mathbb{R}^{n \times N}$ is a known measurement matrix, and $\v{y} \in \mathbb{R}^{n \times 1}$ is the measured signal with $n \ll N$. In the CS context, $n$ is the number of samples sensed while $N$ is the number of samples in the original signal. We take $||\hat{\v{x}}||_0$ to denote the $\ell_0$ pseudo norm from \cite{donoho2006b}, i.e. the number of non-zero entries in $\hat{\v{x}}$. Solving the combinatorial problem in (\ref{eq:compressive_sensing-general_problem}) by an exhaustive search is generally infeasible.

One feasible approach in reconstructing the signal is to relax the problem in (\ref{eq:compressive_sensing-general_problem}) by substituting the $\ell_1$ norm for the $\ell_0$ making the problem a linear program (LP) \cite{donoho2004}. Another feasible approach is taken by the family of so-called iterative greedy algorithms. In these, the problem in (\ref{eq:compressive_sensing-general_problem}) is reversed by minimising the residual of the energy of $||\v{y} - \m{A} \hat{\v{x}}||_2^2$ subject to some sparsity enforcing constraint. Abstractly, the greedy algorithms can be separated into two classes \cite{maleki2010}: 1) Simple one stage algorithms which use a single greedy step in each iteration. Examples are Matching Pursuit (MP) \cite{mallat1993} and Iterative Hard Thresholding (IHT) \cite{blumensath2009c}. 2) Composite two stage algorithms which combine a greedy step with a refinement step in each iteration. Examples are Orthogonal Matching Pursuit (OMP) \cite{tropp2007} and CoSaMP \cite{needell2009a}. The main advantage of the greedy algorithms over the $\ell_1$ approach is that they are computationally less complex \cite{dai2009} and require less computation time than state-of-the-art LP solvers \cite{blumensath2010}.

In addition to the computation time, a measure of the reconstruction quality must be considered. Recently, the measure of phase transition \cite{tanner2010} has become a standard way to specify reconstruction capabilities, see e.g. \cite{maleki2010}, \cite{donoho2010b}, \cite{jain2011}, \cite{sturm2011}. Phase transitions evaluate the probability of successful reconstruction versus the indeterminacy of the constraints $\v{y} = \m{A} \hat{\v{x}}$ and the true sparsity of $\hat{\v{x}}$. In general, the main advantage of the $\ell_1$ approach over the greedy algorithms is that it is superior in terms of phase transition \cite{maleki2010}.

In search of a fast algorithm with a phase transition similar to that of the $\ell_1$ approach, it has been proposed to solve (\ref{eq:compressive_sensing-general_problem}) by approximating the $\ell_0$ norm with a continuous function \cite{mohimani2009}. The resulting smoothed $\ell_0$ norm (SL0) algorithm has a better phase transition than the greedy algorithms while requiring considerably less computation time than the state-of-the-art LP solvers. In this paper, we show that a few key parameters must be carefully selected and knowledge of the indeterminacy exploited to fully unleash the potential of SL0. We provide a set of empirically determined recommended parameters for a modified SL0 algorithm that may dramatically improve its phase transition. Through extensive simulations, the claim of superiority of the recommended parameters is supported. Finally, we discuss implementation strategies that speed up the algorithm by exploiting knowledge of the indeterminacy.

The paper is organised as follows. In Section \ref{sec:SL0}, we restate the SL0 algorithm and present the proposed algorithm. Implementations of SL0 that yield reduced computation time are discussed in Section \ref{sec:fast_implementations}. Section \ref{sec:simulation_framework} describes the setup used for simulations while Section \ref{sec:results} provides the simulation results. A discussion of the results is given in Section \ref{sec:discussion}. Finally, conclusions are stated in Section \ref{sec:conclusions}.

\section{Smoothed $\ell_{0}$ Norm} \label{sec:SL0} \noindent SL0 attempts to solve the problem in (\ref{eq:compressive_sensing-general_problem}) by approximating the $\ell_0$ norm with a continuous function. Consider the continuous Gaussian function $f_{\sigma}$ with the parameter $\sigma$:

\begin{align} f_{\sigma}(x) &= \exp{-x^2/(2 \g \sigma^2)}, \quad x \in \mathbb{R},\; \sigma \in \mathbb{R}_+ \label{eq:gaussian} \end{align}

The parameter $\sigma$ may be used to control the accuracy with which $f_{\sigma}$ approximates the Kronecker delta. In mathematical terms, we have \cite{mohimani2009}:

\begin{align} \lim_{\sigma \rightarrow 0}f_{\sigma}(x) &= \left\{ \begin{array}{ll} 1, & x = 0 \\ 0, & x \neq 0 \end{array} \right. \\ f_{\sigma}(x) &\approx \left\{ \begin{array}{ll} 1, & \abs{x} \ll \sigma \\ 0, & \abs{x} \gg \sigma \end{array} \right. \end{align}

Define the continuous multivariate function $g$ as:

\begin{align} g(\v{x}) \triangleq \sum_{i = 1}^{N} f_{\sigma}(x_i), \quad\v{x} \in \mathbb{R}^{N \times 1} \end{align}

Since the number of entries in $\hat{\v{x}}$ is $N$ and the function $g$ is an indicator of the number of zero-entries in $\hat{\v{x}}$, the $\ell_0$ norm of the reconstructed vector $\hat{\v{x}}$ is approximated by:

\begin{align} ||\hat{\v{x}}||_0 \approx N - g(\hat{\v{x}}) \end{align}

Substituting this approximation into (\ref{eq:compressive_sensing-general_problem}) yields the problem:

\begin{align} \begin{aligned} &\text{minimise} && N - g(\hat{\v{x}}) \\ &\text{subject to} && \v{y} = \m{A} \hat{\v{x}} \end{aligned} \label{eq:compressive_sensing-sl0_problem} \end{align}

The approach is then to solve the problem in (\ref{eq:compressive_sensing-sl0_problem}) for a decreasing sequence of $\sigma$'s. The underlying thought is to select a $\sigma$ which ensures that the initial solution is in the subset of $\mathbb{R}^{N \times 1}$ over which the approximation is convex and gradually increase the accuracy of the approximation. By careful selection of the sequence of $\sigma$'s, (hopefully) non-convexity and thereby local minima are avoided. In the SL0 algorithm stated below, we let $\m{A}^{\dagger}$ denote the Moore-Penrose pseudo-inverse of the matrix $\m{A}$ and let $\v{x} \circ \v{y}$ denote the Hadamard product (entry wise multiplication) of the vectors $\v{x}$ and $\v{y}$. Furthermore, we let $\exp{\v{x}} = [\exp{x_1} \dots \exp{x_N}]^T$ and $\max\abs{\v{x}} = \max\{\abs{x_1}, \dots, \abs{x_N}\}$.

\vspace{0.4em} \noindent \textbf{SL0 algorithm:} \begin{algorithmic}[1] \STATE \textbf{initialise:} $\sigma_|up| = 0.5$, $\sigma_|min| = 0.01$, $\mu = 1$, $L = 3$ \STATE $\hat{\v{x}} \algarrow \m{A}^{\dagger}\v{y}$ \STATE $\sigma \algarrow 2 \g \max\abs{\hat{\v{x}}}$ \vspace{0.2cm} \WHILE{$\sigma > \sigma_|min|$} \FOR{$i = 1 \dots L$} \STATE $\v{d} \algarrow \hat{\v{x}} \circ \exp{-(\hat{\v{x}} \circ \hat{\v{x}}) / (2 \g \sigma^2)}$ \label{alg:sl0-sl0_basic_algorithm-d} \STATE $\hat{\v{x}} \algarrow \hat{\v{x}} - \mu \g \v{d}$ \label{alg:sl0-sl0_basic_algorithm-x1} \STATE $\hat{\v{x}} \algarrow \hat{\v{x}} - \m{A}^{\dagger}(\m{A}\hat{\v{x}} - \v{y})$ \label{alg:sl0-sl0_basic_algorithm-x2} \ENDFOR \STATE $\sigma \algarrow \sigma \g \sigma_|up|$ \ENDWHILE \end{algorithmic} \vspace{0.4em}

\noindent For each $\sigma$, the problem in (\ref{eq:compressive_sensing-sl0_problem}) is solved by repeatedly taking an unconstrained gradient descent step in line \ref{alg:sl0-sl0_basic_algorithm-x1} ($\v{d}$ is a normalised gradient of $g$) and projecting $\hat{\v{x}}$ back onto the feasible set in line \ref{alg:sl0-sl0_basic_algorithm-x2}. Substituting line \ref{alg:sl0-sl0_basic_algorithm-x1} into line \ref{alg:sl0-sl0_basic_algorithm-x2} results in the expression:

\begin{align} \hat{\v{x}} &\algarrow \hat{\v{x}} - \mu \g (\m{I} - \m{A}^{\dagger}\m{A})\v{d} \label{eq:sl0-x_update} \end{align}

where $\m{I} \in \mathbb{R}^{N \times N}$ is the identity matrix. The matrix $\m{I} - \m{A}^{\dagger}\m{A}$ is a projector onto the null space of $\m{A}$. Consequently, as pointed out in \cite{cui2010}, the unconstrained gradient descent step followed by projection back onto the feasible set is equivalent to a direct gradient descent step on the feasible set. The reason is that the projector $\m{I} - \m{A}^{\dagger}\m{A}$ restricts the gradient descent step to be on the feasible set. The parameter $\sigma_|up|$ is the constant (typically chosen in the interval $[0.5; 1[$) in the geometric sequence of $\sigma$'s while $\sigma_|min|$ relates to the final value used for $\sigma$ and hence this controls the quality of the reconstruction. The optimal choice of $\sigma_|up|$ and $\sigma_|min|$ is problem dependent while an initial $\sigma = 2 \g \max\abs{\hat{\v{x}}}$ as well as $\mu = 1$ and $L = 3$ are problem independent recommendations given in \cite{mohimani2009}.

\subsection{Improving the Phase Transition} \label{sec:improving_phase_transition}

\noindent A comprehensive proof of the convergence of the SL0 algorithm exists for a specific set of parameters provided that an Asymmetric Resticted Isometry Property (ARIP) constraint is satisfied \cite{mohimani2010}. The authors of the proof conclude that though theoretically satisfactory, the ARIP constraint leads to an unnecessarily pessimistic choice of parameter values. Motivated by this conclusion, we have carried out an extensive empirical analysis with the objective of finding the optimal parameter values. This analysis has shown that the standard step-size $\mu = 1$ and iteration number $L = 3$ provided in \cite{mohimani2009} result in a suboptimal phase transition. Instead, we suggest a modification to the SL0 algorithm which may greatly improve the phase transition.

Our initial observation is that the SL0 algorithm presented here is outperformed by the greedy IHT algorithm described in \cite{maleki2010} in terms of phase transition for indeterminacy $\delta = n / N < 0.8$. By introducing an update of the $L$ parameter, the phase transition of the SL0 algorithm can be improved to lie between that of the greedy IHT algorithm and that of the $\ell_1$ approach. Specifically, we obtained this improvement by multiplying $L$ with a factor $L_|up| \geq 1.5$ after each update of $\sigma$. By carefully selecting sequences of $\mu$'s and $\sigma$'s, the phase transition of the SL0 algorithm can be even further improved to consistently lie on or above that of the $\ell_1$ approach. In brief, we chose the step-size in the order of $10^{-3}$ for the first few $\sigma$'s and in the order of $1$ for the last $\sigma$'s. Also, we chose the initial $\sigma$ based on knowledge of the indeterminacy $\delta$ to improve the phase transition across all $\delta$.

Based on our experimental results, we propose the following strategy. We choose a step-size of $\mu = 0.001$ for the first three $\sigma$'s and a step-size of $\mu = 1.4$ for the last $\sigma$'s. We use a sequence of $\bm{\mu} = [0.001, 0.001, 0.001, 0.05, 0.06, 1.4, \dots, 1.4]\T$ where the number of entries equals the number of $\sigma$'s used. Furthermore, an inversely proportional relation between $\delta$ and the initial value of $\sigma$ yielded the most promising phase transition. Specifically, we choose an initial $\sigma = 1 / (2.75 \g \delta) \g \max\abs{\hat{\v{x}}}$ and combine this choice with a $\sigma_|up| = 0.7$. Finally, a gradually increasing $L$ for decreasing $\sigma$ still provides an improvement for the updated parameter choices. Here, we choose a geometric sequence starting with $L = 2$ and increasing by a factor of $L_|up| = 1.9$ for each update of $\sigma$.

With an increased value of $\sigma_|up|$ and gradually increasing values of $L$, the computation time is bound to increase. This effect can, however, be counteracted by introducing a stopping criterion in the inner loop of the SL0 algorithm. Therefore, we choose to terminate the inner loop when the relative change $||\hat{\v{x}} - \hat{\v{x}}_|prev|||_2$ falls below $\sigma \g \epsilon$ where $\epsilon = 0.01$. Generally, this measure has proved to be a good indicator of convergence and significantly reduced the average number of iterations taken in the inner loop. We now propose the smoothed $\ell_0$ norm algorithm with modified step-size (SL0 MSS) which incorporates all of the above findings.

\vspace{0.4em} \noindent \textbf{SL0 MSS algorithm:} \begin{algorithmic}[1] \STATE \textbf{initialise:} $\sigma_|up| = 0.7$, $\sigma_|min| = 0.01$, $L = 2.0$, \\ \hspace{1.62cm}$L_|up| = 1.9$, $\epsilon = 10^{-2}$, $k = 1$ \STATE $\hat{\v{x}} \algarrow \m{A}^{\dagger}\v{y}$ \label{alg:SL0_mod-LS_init} \STATE $\bm{\mu} \algarrow [0.001, 0.001, 0.001, 0.05, 0.06, 1.4, ... , 1.4]\T$ \STATE $\sigma \algarrow 1/(2.75 \g \delta) \g \max\abs{\hat{\v{x}}}$ \vspace{0.2cm} \WHILE{$\sigma > \sigma_|min|$} \STATE $\hat{\v{x}}_|prev| \algarrow \v{0}$ \STATE $i = 0$ \WHILE{$||\hat{\v{x}} - \hat{\v{x}}_|prev|||_2 > \sigma \g \epsilon$ \AND $i < L$} \label{alg:SL0_mod-inner_loop} \STATE $\hat{\v{x}}_|prev| \algarrow \hat{\v{x}}$ \STATE $\v{d} \algarrow \hat{\v{x}} \circ \exp{-(\hat{\v{x}} \circ \hat{\v{x}}) / (2 \g \sigma^2)}$ \STATE $\hat{\v{x}} \algarrow \hat{\v{x}} - \mu_{k} \g (\m{I} - \m{A}^{\dagger}\m{A})\T\v{d}$ \label{alg:SL0_mod-x_update} \STATE $i \algarrow i + 1$ \ENDWHILE \STATE $\sigma \algarrow \sigma \g \sigma_|up|;\,\,\,$ $L \algarrow L \g L_|up|;\,\,\,$ $k \algarrow k + 1;\,\,\,$ \ENDWHILE \end{algorithmic} \vspace{0.4em}

\section{Implementations of SL0} \label{sec:fast_implementations} \noindent Through experiments, we have found the most time consuming parts of the SL0 MSS algorithm to be the computation of $\m{A}^{\dagger}$ and the matrix-vector products in the gradient descent step. We now provide a strategy for exploiting the known indeterminacy $\delta = n / N$ to reduce the number of required computations.

Consider the full QR decomposition of $\m{A}\T$ \cite{boyd2004}:

\begin{align} \m{A}\T &= [\m{Q}_1 \: \m{Q}_2] \begin{bmatrix} \m{R} \\ \m{0} \end{bmatrix} \end{align}

where $\m{Q}_1 \in \mathbb{R}^{N \times n}$ and $\m{Q}_2 \in \mathbb{R}^{N \times (N-n)}$ form bases for the row space and null space of $\m{A}$, respectively and $\m{R} \in \mathbb{R}^{n \times n}$ is upper triangular. Note how the dimensions of $\m{Q}_1$, $\m{Q}_2$, and $\m{R}$ change with $\delta$. From the QR decomposition, the pseudo-inverse $\m{A}^{\dagger}$ can be found:

\begin{align} \m{A}^{\dagger} &= \m{Q}_1\m{R}\MT \end{align}

Three equivalent expressions for the projector that projects onto the null space of $\m{A}$ are then:

\begin{align} \m{I} - \m{A}^{\dagger}\m{A} = \m{I} - \m{Q}_1\m{Q}_1\T = \m{Q}_2\m{Q}_2\T \end{align}

We now propose the following scheme for an implementation of the SL0 MSS algorithm. When $\delta \leq 0.5$, use the projector $\m{A}^{\dagger}$ and split the update in line \ref{alg:SL0_mod-x_update} in an infeasible gradient descent step followed by a projection onto the feasible set:

\begin{align} \hat{\v{x}} &\algarrow \hat{\v{x}} - \mu \g \v{d} \label{eq:SL0_mod_MSSI_1} \\ \hat{\v{x}} &\algarrow \hat{\v{x}} - \m{A}^{\dagger}(\m{A}\hat{\v{x}} - \v{y}) \label{eq:SL0_mod_MSSI_2} \end{align}

When $\delta > 0.5$, use the projector $\m{Q}_2\m{Q}_2\T$ and split the update in line \ref{alg:SL0_mod-x_update} in two steps:

\begin{align} \v{v} &\algarrow \m{Q}_2\T\v{d} \label{eq:QR_methods-split1} \\ \hat{\v{x}} &\algarrow \hat{\v{x}} - \mu_{k} \g \m{Q}_2\v{v} \label{eq:QR_methods-split2} \end{align}

To appreciate the split procedure, note that fewer arithmetic operations are needed in computing the matrix-vector product $(\m{Q}_2 \m{Q}_2\T) \v{d}$ this way when:

\begin{align} &\mathrm{Flops}\{(\m{Q}_2 \m{Q}_2\T) \v{d}\} > \nonumber \\ &\quad\mathrm{Flops}\{\m{Q}_2\T\v{d}\} + \mathrm{Flops}\{\m{Q}_2\v{v}\} \intertext{which implies that:} &N \g (2N - 1) > \nonumber \\ &\quad(N-n) \g (2N - 1) + N \g (2\g(N-n) - 1) \intertext{or using an approximation:} &\qquad\qquad\quad n \gtrapprox N / 2 \: \leftrightarrow \: \delta \gtrapprox 1/2 \end{align}

where $\mathrm{Flops}\{x\}$ denotes the number of floating point operations in the computation of the expression $x$. The split procedure does not require the explicit forming of $\m{A}^{\dagger}$ since the initial least squares (least norm) solution $\hat{\v{x}} = \m{A}^{\dagger}\v{y}$ in line \ref{alg:SL0_mod-LS_init} may be efficiently computed by solving the following system of equations by substitution (since $\v{R}$ is triangular):

\begin{align} \v{y} = \m{R}\T\v{u} \label{eq:sl0-finding_matrices-implicit_ls1} \end{align}

followed by computing:

\begin{align} \hat{\v{x}} = \m{Q}_1\v{u} \label{eq:sl0-finding_matrices-implicit_ls2} \end{align}

Avoiding the explicit forming of $\m{A}^{\dagger}$ eliminates the need for computing the inverse of $\m{R}\T$ which becomes progressively more computationally expensive as $\delta \rightarrow 1$. On the other hand, only the reduced QR decomposition $\m{A}\T = \m{Q}_{1}\m{R}$ is needed when $\m{A}^{\dagger}$ is explicitly formed. The reduced QR decomposition becomes progressively less computationally expensive than the full QR decomposition as $\delta \rightarrow 0$. These two observations motivate the change of method at $\delta = 1/2$.

\section{Simulation Framework} \label{sec:simulation_framework} \noindent We evaluate the reconstruction capabilities of an algorithm by use of the phase transition measure \cite{tanner2010} which provides the following framework. Let $k$ denote the number of non-zero entries in the true signal $\v{x}$. Define the measure of indeterminacy $\delta = n / N$ and the generalised measure of sparsity (density) $\rho = k / n$. Given a success criterion, the success rate is then evaluated on the phase space $(\delta, \rho) \in [0, 1]^2$. In general, reconstruction is easier for larger $\delta$ and smaller $\rho$ and becomes increasingly difficult when decreasing $\delta$ and increasing $\rho$. Somewhere in-between, the phase transition curve separates the phase space into a phase where reconstruction is likely and a phase where it is unlikely. This phase transition curve is continuous in $\delta$ for fixed $N$. Obviously, it is desirable to have a phase transition curve which is as close to $\rho = 1$ as possible.

Different suites of problems, i.e. different combinations of ensembles of $\m{A}$ and $\v{x}$ generally result in different phase transitions. Choosing $\m{A}$ from the Uniform Spherical Ensemble (USE) and the non-zero entries in $\v{x}$ from the Rademacher distributed generally yields the most difficult problem suite in terms of obtaining good phase transitions \cite{maleki2010}. In the simulations, we consider this problem suite along with a the problem suite where $\m{A}$ is chosen from the USE and the non-zero entries in $\v{x}$ are chosen from the zero mean, unit variance Gaussian ensemble. In \cite{donoho2009a}, it is shown that the probability of reconstruction versus $\rho$ for fixed $N$ and $\delta$ can be modelled accurately by logistic regression for a specific set of algorithms. Logistic regression is used more generally in \cite{maleki2010} to determine the location of the phase transition curve. We adopt the logistic regression approach to estimate the location of the phase transition curve and fix $N = 800$ as proposed in \cite{maleki2010}. We then attempt reconstruction on a uniform grid $(\delta, \rho)$ in the phase space specified by:

\begin{align} \delta \in \{0.025, 0.05, \dots, 1.00\} \\ \rho \in \{0.01, 0.02, \dots, 1.00\} \end{align}

For each point in the grid, we do 10 Monte Carlo simulations where each simulation features a new draw of $\m{A}$ and $\v{x}$. From \cite{maleki2010} we adopt that an attempted reconstruction is considered successful when:

\begin{align} \frac{||\hat{\v{x}} - \v{x}||_2^2}{||\v{x}||_2^2} < 10^{-4} \end{align}

where $\hat{\v{x}}$ and $\v{x}$ are the reconstructed and true signal, respectively. If the criterion is not met, the attempted reconstruction is considered unsuccessful, i.e., the attempted reconstruction cannot be considered indeterminate.

To evaluate the required computation time for the different algorithms, we measure the absolute time spent on reconstruction when it succeeds. The problem suite formed by choosing $\m{A}$ from the USE combined with Rademacher distributed non-zero entries in $\v{x}$ is used in this test. A uniform grid $(\delta, \rho)$ in the phase space is formed by:

\begin{align} \delta \in \{0.1, 0.2, \dots, 1.0\} \\ \rho \in \{0.1, 0.2, \dots, 1.0\} \end{align}

An algorithm is tested on all points in the grid that are at least 0.025 below its empirically determined phase transition (measured on the $\rho$-axis). The time spent on reconstruction for the successful part of 10 Monte Carlo simulations in each point is then averaged. Considered problem sizes are:

\begin{align} N \in \{800, 1600, 3200, 6400, 12800\} \end{align}

The simulations have been conducted on an Intel Core i7 970 6-core \SI{3.2}{\giga\hertz} based PC with \SI{24}{\gibi\byte} DDR3 RAM. The OS used is 64-bit Ubuntu 12.04 LTS Linux and the Enthought Python Distribution (EPD) 7.2-2 (64-bit). All simulations are carried out in double precision.

To validate the results obtained from our simulation framework, we have simulated the Iterative Hard Thresholding algorithm presented in \cite{maleki2010}. The phase transition obtained in our simulation framework has then been compared with the phase transition obtained in the simulation framework of \cite{maleki2010}. Due to the non-deterministic nature of the Monte Carlo simulations used in both simulation frameworks, the two phase transitions will inevitably differ slightly. However, we have observed that they are almost identical and therefore concluded, that our simulation framework works as intended.

\section{Experimental Results} \label{sec:results} \noindent Four algorithms have been simulated: 1) SL0 STD which is the SL0 algorithm presented in Section \ref{sec:SL0}. 2) SL0 MIN which is the same algorithm except it is modified such that $L$ is multiplied by $1.7$ each time $\sigma$ is decreased. 3) SL0 MSS which is the algorithm presented in Section \ref{sec:improving_phase_transition}. 4) IHT which is the Iterative Hard Thresholding algorithm described in \cite{maleki2010}. In the case of the SL0 MSS algorithm, two implementations have been simulated: The SL0 MSS\rom{1} implementation based on (\ref{eq:SL0_mod_MSSI_1}) and (\ref{eq:SL0_mod_MSSI_2}) and the SL0 MSS\rom{2} implementation based on (\ref{eq:QR_methods-split1}) and (\ref{eq:QR_methods-split2}).

\begin{figure*}[t] \centering \begin{minipage}{0.49\textwidth} \centering \includegraphics[width=1.0\textwidth]{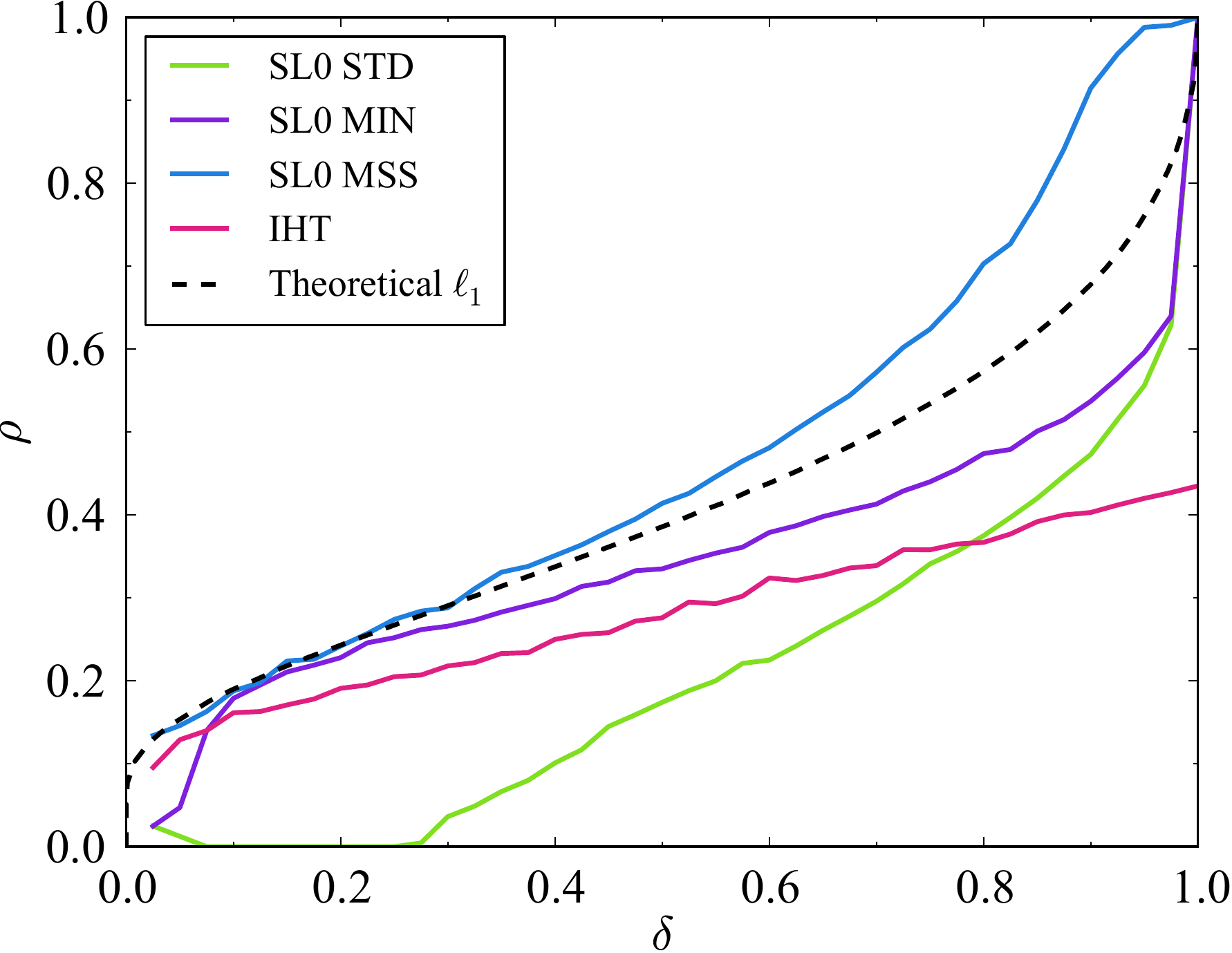} \caption{Phase transitions for the problem suite: $\m{A}$ from the USE, $\v{x}$ with non-zero Rademacher entries. The theoretical $\ell_1$ curve is from \cite{theoretical_l1}.} \label{fig:results-rademacher-phase} \end{minipage} \hfill \begin{minipage}{0.49\textwidth} \centering \includegraphics[width=1.0\textwidth]{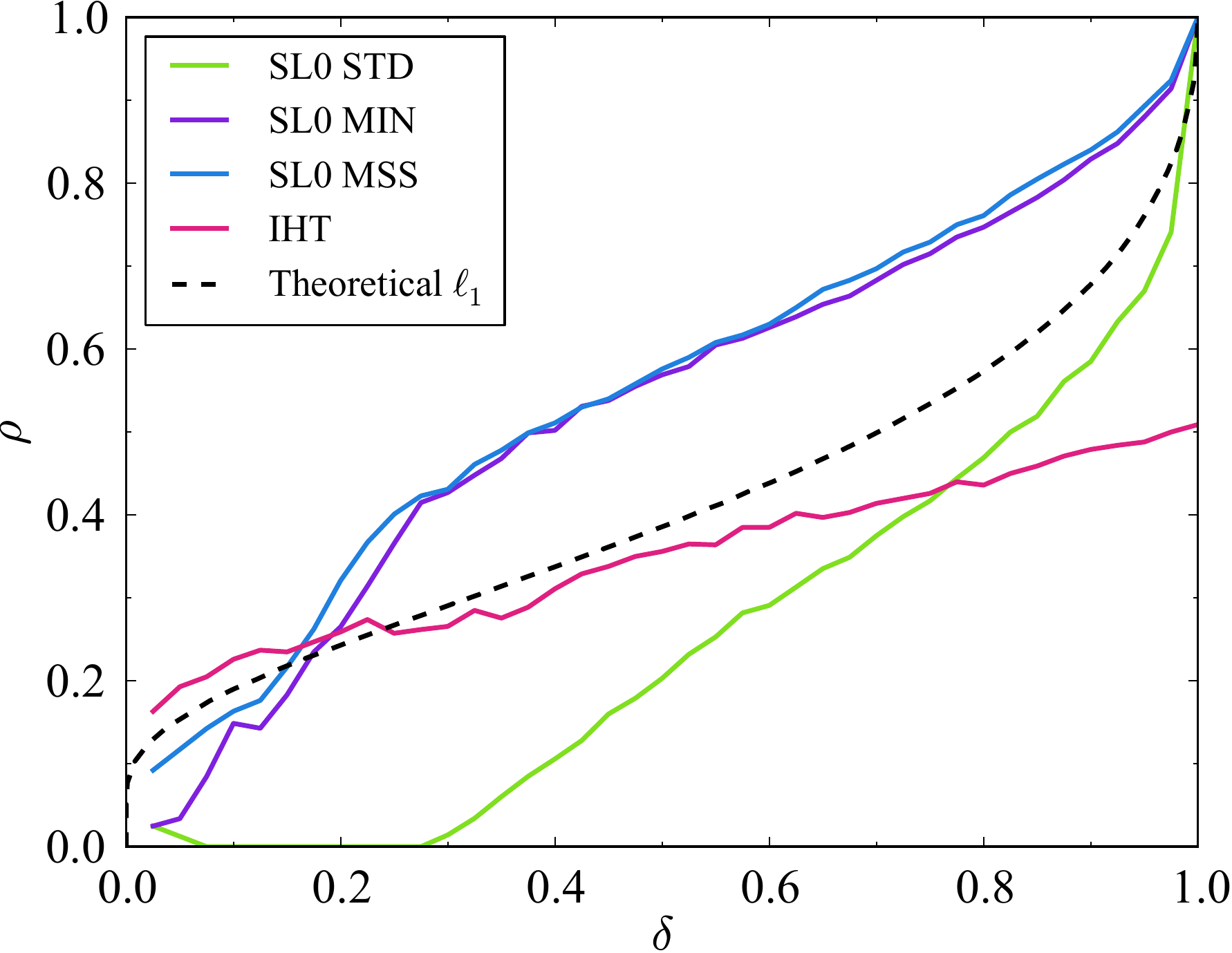} \caption{Phase transitions for the problem suite: $\m{A}$ from the USE, $\v{x}$ with non-zero zero mean, unit variance Gaussian entries. The theoretical $\ell_1$ curve is from \cite{theoretical_l1}.} \label{fig:results-gaussian-phase} \end{minipage} \end{figure*}

\begin{figure*}[t] \centering \begin{minipage}{0.49\textwidth} \centering \includegraphics[width=1.0\textwidth]{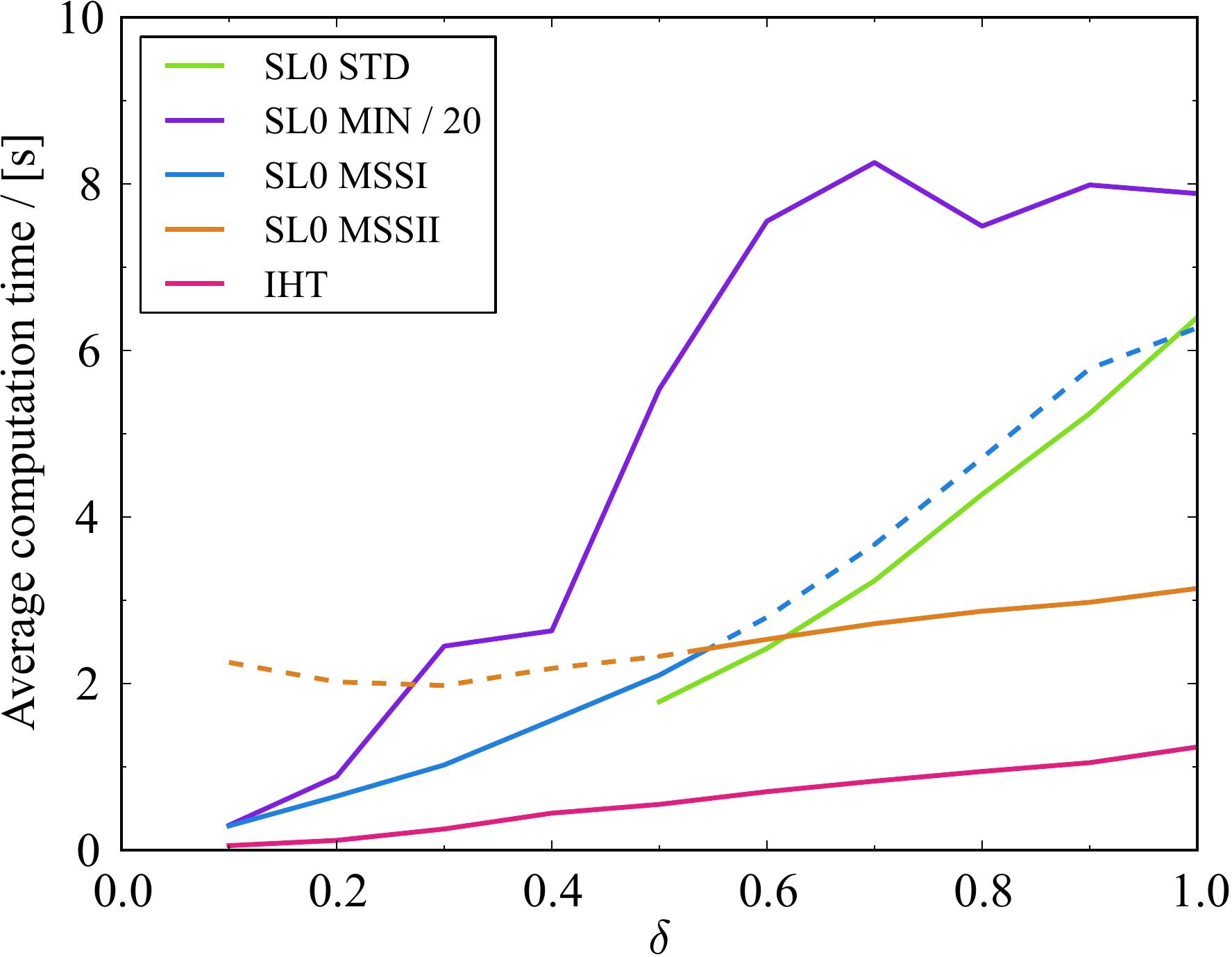} \caption{Measured average computation times versus indeterminacy $\delta = n/N$ for fixed $N=3200$.} \label{fig:results-computation_times} \end{minipage} \hfill \begin{minipage}{0.49\textwidth} \centering \includegraphics[width=1.0\textwidth]{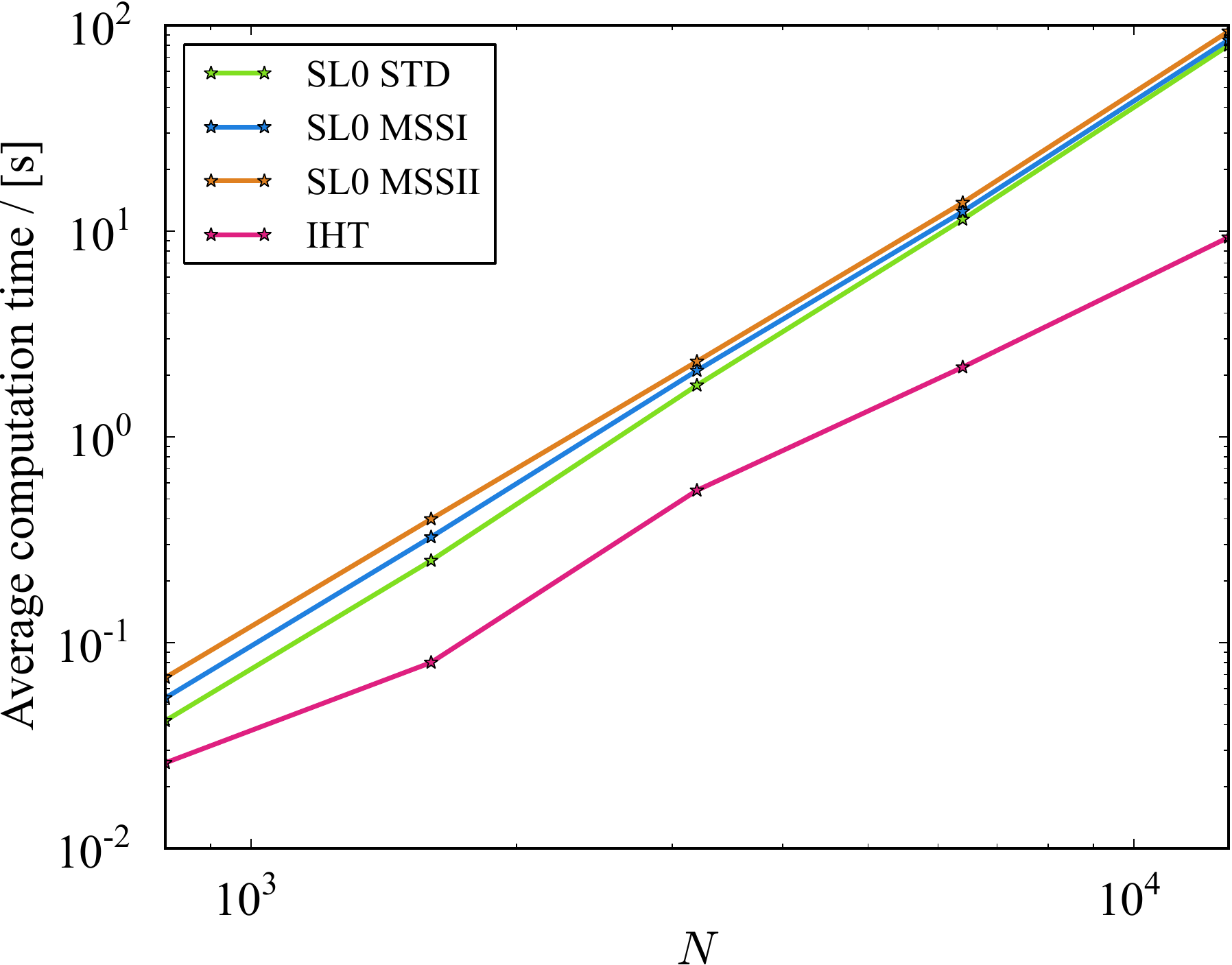} \caption{Measured average computation times versus problem size $N$ for fixed $\delta = 1/2$.} \label{fig:results-scaling} \end{minipage} \end{figure*}

The experimental results are presented in Figure \ref{fig:results-rademacher-phase}, \ref{fig:results-gaussian-phase}, \ref{fig:results-computation_times}, and \ref{fig:results-scaling}. Figure \ref{fig:results-rademacher-phase} shows the phase transitions for Rademacher distributed non-zero entries in $\v{x}$ while Figure \ref{fig:results-gaussian-phase} shows the phase transitions for zero mean, unit variance Gaussian non-zero entries in $\v{x}$. In both figures, the theoretical $\ell_1$ curve from \cite{theoretical_l1} is included for reference. Figure \ref{fig:results-computation_times} shows the measured average computation times versus indeterminacy $\delta = n / N$ and Figure \ref{fig:results-scaling} shows the measured average computation times versus problem size $N$. Note the abrupt ending of the SL0 STD curve in Figure \ref{fig:results-computation_times} which is due to failure of reconstruction in the tested grid for $\delta < 1 / 2$. Also, note that the measured computation times of the SL0 MIN implementation have been divided by 20.

In summary, SL0 MSS shows the best phase transition among the tested algorithms for both Rademacher and Gaussian non-zero entries in $\v{x}$. In large portions of the phase space it even surpasses the theoretical $\ell_1$ curve. The only exception is in the Gaussian case for $\delta < 0.2$ where IHT shows better phase transition. Regarding computation time, IHT is faster than the SL0 approaches among which SL0 MIN is consistently more than 20 times slower than SL0 MSS. Furthermore, IHT scales slightly better with problem size $N$ than the SL0 approaches.

\section{Discussion} \label{sec:discussion} \noindent For Rademacher non-zero entries in $\v{x}$, Figure \ref{fig:results-rademacher-phase} reveals that SL0 STD, SL0 MIN, and IHT are by far outperformed by SL0 MSS in terms of phase transistion. Even the theoretical $\ell_1$ curve is surpassed by SL0 MSS at around $\delta > 0.3$. For $\delta < 0.3$, SL0 MSS shows the same phase transition as the theoretical $\ell_1$ curve. The curve for SL0 MIN is a clear example of the improvement in phase transition obtainable using more iterations in the inner loop of the SL0 algorithm. However, SL0 MSS further improves on this, especially for $\delta < 0.1$ and $\delta > 0.3$.

The results in Figure \ref{fig:results-computation_times} settle that SL0 does indeed require more computation time than IHT. IHT is around two to four times faster (depending on $\delta$) than the fastest SL0 implementation for a problem size of $N=3200$. The important thing to note though, is that SL0 provides a trade-off between phase transition and computation time. The price paid in computation time for using a lot of iterations to get better phase transition is clear from the SL0 MIN curve. This is, however, not the case for SL0 MSS, which requires less than or about the same computation time as SL0 STD depending on $\delta$. Thus, a much better phase transition is obtained using largely the same computation time in going from SL0 STD to SL0 MSS. To obtain such a result, it is necessary to switch from SL0 MSS\rom{1} to SL0 MSS\rom{2} at around $\delta = 1/2$.

An assessment of the scaling of average computation time with problem size $N$ reveals that all three SL0 algorithms seem to scale in an equivalent way. IHT scales better than SL0 and hence requires relatively less computation time as the problem size $N$ increases. The scaling depicted in Figure \ref{fig:results-scaling} is for $\delta = 1/2$ which provides a rough average computation time across all values of $\delta$ as can be seen from Figure \ref{fig:results-computation_times}.

The parameters for SL0 MSS stated in Section \ref{sec:improving_phase_transition} are (locally) optimal in terms of phase transition for Rademacher non-zero entries in $\v{x}$. Gaussian non-zero entries in $\v{x}$ are known to be in favour of greedy algorithms \cite{sturm2011}, which is also the case for IHT in our simulations, especially for $\delta < 0.2$ where the IHT curve surpasses the theoretical $\ell_1$ curve. For $\delta > 0.2$, SL0 MIN and SL0 MSS demonstrate the best phase transition among the shown algorithms. SL0 MIN and SL0 MSS phase transitions are about the same, though. Comparing our results for SL0 MSS in Figure \ref{fig:results-gaussian-phase} with the ones given for SL0 in Figure 6 in \cite{sturm2011} shows about the same phase transition. The slightly better phase transition for $\delta < 0.2$ in \cite{sturm2011} may be due to the SL0 MSS parameters not necessarily being optimal for Gaussian non-zero entries in $\v{x}$.

Although the above simulations are quite encouraging, they are based on an empirically tuned algorithm. Thus, to reach a final verdict of the success of SL0 MSS, the validity of the simulation results must be exhaustively studied for a broader set of problem suites. Alternatively, more sound mathematical proofs must be presented.

\section{Conclusions} \label{sec:conclusions} \noindent We have proposed a new compressive sensing reconstruction algorithm named SL0 MSS based on the smoothed $\ell_0$ norm. It turns out that SL0 phase transitions heavily depend on parameter selection. SL0 MSS attempts to improve on phase transition by exploiting the known indeterminacy $\delta = n/N$ combined with carefully selected parameters. A trade-off between phase transition and computation time is provided by SL0. Improved phase transition has been measured for SL0 MSS compared to standard SL0 while maintaining the same computation time.

\label{biblograph} \end{document}